# Flattening van der Waals heterostructure interfaces by local thermal treatment

Justin Boddison-Chouinard[1]*, Samantha Scarfe[1]*, K. Watanabe[2], T. Taniguchi[2], Adina Luican-Mayer[1]†

[1]Department of Physics, University of Ottawa, Ottawa, K1N 6N5, Canada

[2]National Institute for Materials Science, 1-1 Namiki, Tsukuba 305-0044, Japan

*equal contributions

†luican-mayer@uottawa.ca

Fabrication of custom-built heterostructures based on stacked 2D materials provides an effective method to controllably tune electronic and optical properties. To that end, optimizing fabrication techniques for building these heterostructures is imperative. A common challenge in layer-by-layer assembly of 2D materials is the formation of bubbles at the atomically thin interfaces. We propose a technique for addressing this issue by removing the bubbles formed at the heterostructure interface in a custom-defined area using the heat generated by a laser, equipped with raster scanning capabilities. We demonstrate that the density of bubbles formed at graphene-$ReS_2$ interfaces can be controllably reduced using this method. We discuss an understanding of the flattening mechanism by considering the interplay of interface thermal conductivities and adhesion energies between two atomically thin 2D materials.

The ability to design and synthesize high-quality materials with minimal contamination and low defect density has historically resulted in scientific and technological breakthroughs. The family of layered van der Waals 2D crystals represent a promising platform for developing new classes of custom-tailored materials [1]. It comprises a wide range of electronic properties such as insulating (hBN), semiconducting ($MoS_2$, $WS_2$), metallic transitioning into correlated states of matter when cooled down: charge density order (1T-$TaS_2$, $TiSe_2$) or superconductivity ($NbSe_2$, 1H-$TaS_2$)[2]. The ability to controllably stack their atomically thin layers can lead to the realization of original materials with properties previously inaccessible[3]. Experimentally, a key step in achieving this goal is to ensure pristine interfaces between layers. One of the typical difficulties in layer-by-layer assembly of 2D materials is the formation of bubbles at the atomically thin interfaces, which introduce inhomogeneities in the device's electronic properties [4-8].

Van der Waals heterostructures are commonly assembled by a variety of dry mechanical stacking techniques [9,10]. Such methods generally involve bringing a bottom-2D material into contact with a stamp supporting a top-2D material and, frequently, environmental contaminants are trapped at the interface during this process. These environmental contaminants aggregate into inhomogeneously distributed blisters (or "bubbles"), hundreds of nm wide and of tens of nanometers tall, as schematically indicated in Figure 1(a).

Bubbles have been observed to be not only a result of the dry stamping method, but also after irradiation with proton beams or chemical etching processes[11]. These bubbles have been extensively studied ranging from proposals on how to minimize their formation [4,9,12-15] to addressing their content and shape/size distribution [16-21]. The atomic force microscopy (AFM) typograph in Figure 1(b) shows the surface morphology of a $MoS_2$/BN heterostructure [10], with bubbles formed at the interface as indicated. These types of nano-blister shapes are similar to those commonly reported in the literature [17,22]. We observe that these blisters are interconnected by narrow channels suggesting that the trapped contaminants might be able to traverse the sample through pressure or heat gradients [11]. While there is no firm consensus on the contents of these bubbles, they are widely believed to be trapped hydrocarbons from air contaminants present in ambient lab conditions [5] or water [4,5,17,18,21]. It was found that from the aspect ratio of the bubbles one can infer the adhesion energy of a specific interface [17,18,20,21]. In Figure 1(c) we plot the variation in aspect-ratios for different heterostructures we measured, in good agreement with those measured across the literature [5,18,23-25].

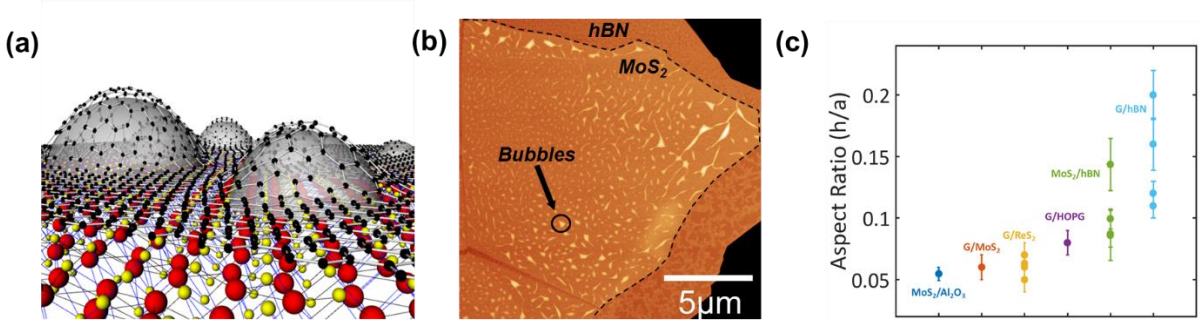

*Figure 1: a) Visual representation of interface contaminants trapped between a graphene film and a transition metal dichalcogenide substrate. b) Atomic force microscope image of $MoS_2$ supported by an hBN substrate. Bubbles are indicated by the green arrow. c) Aspect ratio of measured bubbles formed at the interface of different 2D material stacks.*

Although an interesting playground for studying the elastic properties of atomically thin membranes[17], from a practical perspective, the blisters represent an issue for the device community as they introduce charge inhomogeneities [6] that significantly degrade device performance. Currently, techniques used to eliminate bubbles from heterostructures are either preventive, (careful transfer), non-targeted (thermal treatments), or invasive (pushing the bubbles with an AFM tip in contact mode). Preventive techniques involve slowly laminating the heterostructure at a raised temperature[9]. A common non-targeted technique involves heating the entire sample in a controlled environment at 200°C[12]-300°C [7]; this leads to increased amount of flattened regions as bubbles diffuse under temperature gradients. The drawback of this approach is that it cannot be used when parts of a device might contain crystals that change their structure or are unstable upon heating. Another method relies on the force of an AFM tip to mechanically push bubbles and ultimately flatten the film [15]. Therefore, a targeted and non-invasive technique that addresses the presence of bubbles at the interface of 2D materials is still missing.

Here we present a method that achieves targeted and non-invasive manipulation of bubbles at 2D heterostructure interfaces. By using a localized heat treatment through laser exposure, we combine the diffusive, non-contact nature of a thermal treatment, with the targeted and spatially resolved capabilities of a local probe.

Our method makes use of a 532 nm CW laser with raster scanning capabilities to systematically flatten out 2D heterostructures. Using a Horiba XPlora Plus Raman microscope set-up, we use a 100X objective to focus the laser spot to 0.5µm, and define a path for the laser to collect Raman spectra, collecting a Raman map of a desired heterostructure. We illustrate the use of this method for a heterostructure having a graphene film supported by a thick (30-100nm) $ReS_2$ substrate. Figure 2(a) and Figure 2(b) show the AFM topographic images of a graphene/$ReS_2$ stack before and after laser exposure, respectively. While initially we obtained a dense amount of small bubbles (average height of 15nm), after laser exposure, the bubbles are more sparsely distributed and taller (average height of 80 nm). This is demonstrated by the height histogram of the graphene flake before and after exposure presented in Figure 2(c). We notice that after laser exposure the distribution of heights shifts towards 0 nm, indicating an increase in flattened regions. At the opposite end, the presence of few tall bubbles is only seen in the flake after exposure. The similarities with thermal treatments indicate that the most probable driving mechanism is diffusivity of the bubble contents through heat [11].

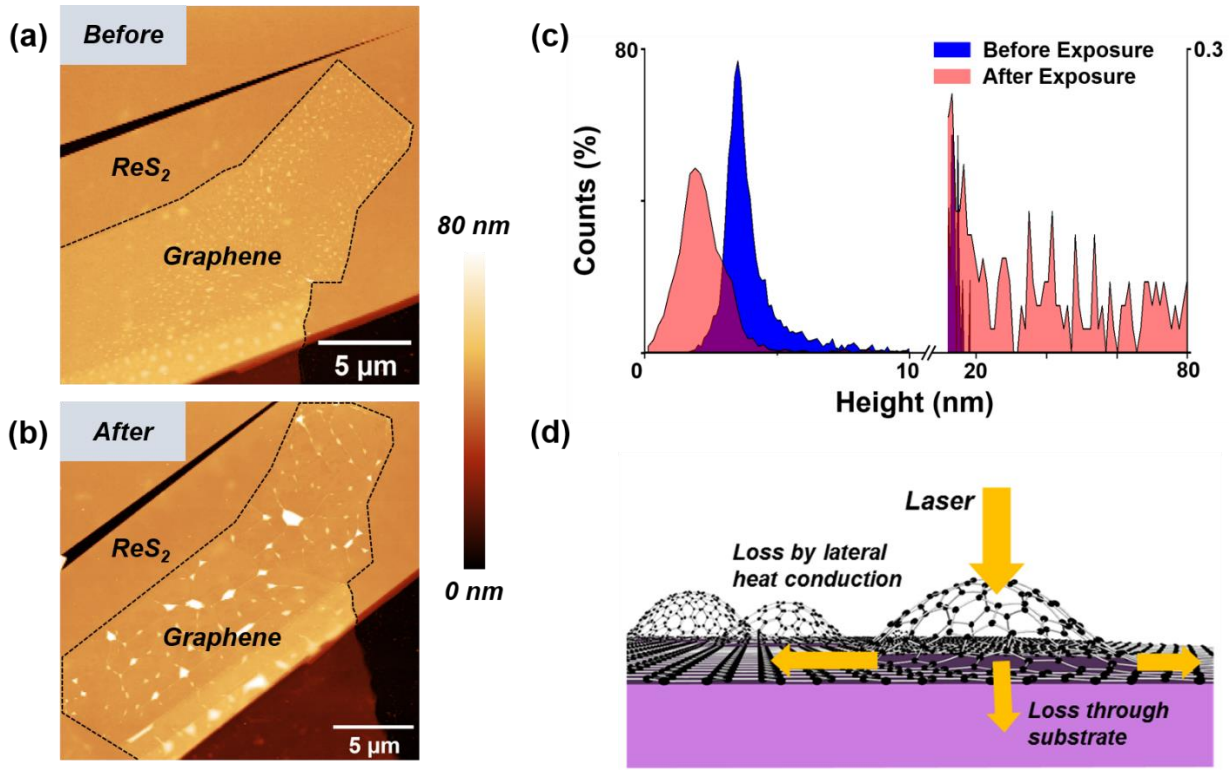

*Figure 2: a) AFM topograph of a graphene flake transferred on a $ReS_2$ substrate. b) AFM of the same graphene flake in (a) after exposure to a rastering laser at 8.7 mW. c) Histogram of AFM pixel heights before and after laser exposure. d) Schematic of the flattening mechanism – interplay of heat loss and surface adhesion.*

We propose the following process: the laser power provides kinetic energy to the bubble content, partly dissipated through the substrate and partly dissipated laterally; the remainder will work against the film-substrate adhesion energy as illustrated in Figure 2(d). Importantly, when the adhesion energy is overcome, the bubble content will be allowed to migrate, leaving behind areas of flat graphene.

By estimating how much energy is needed for the bubble contents to overcome the adhesion energy of a particular heterostructure, we can obtain a minimum temperature for bubble migration as described below.

The adhesion energy per unit area is given by [18]: $E_a = \frac{6E_{2D}h^4}{5a^4} + \gamma_w(cos\theta_t + cos\theta_b)$, where $E_{2D}$ is the in-plane elastic stiffness of the top film ($E_{2D}^{graphene} = 340\ Nm^{-1}$)[18], $h$ is the height of the bubble, $a$ is the bubble radius, $\gamma_w$ is the surface tension of water ($\approx$0.072 J/m$^2$), and $\theta_t$ and $\theta_b$ are the water contact angles of the top crystal and bottom crystal respectively.

We consider the kinetic energy delivered to bubble contents $K_{bubble} = \frac{3}{2}k_B TN$ where $k_B$ is the Boltzmann constant, $T$ is the temperature in the bubble and $N$ are the number of molecules. We assume the contents to be water molecules [18] and we approximate the bubble shape as a spherical cap.

Now we equate $E_{adhesion}$ to $K_{bubble}$ to calculate the minimum bubble migration temperature.

$$T = \frac{4(d^2 - \pi a^2)}{k_B \rho \pi h(3a^2 + h^2)}\left(\frac{6E_{2D}h^4}{5a^4} + \gamma_w(cos\theta_{graphene} + cos\theta_{ReS_2})\right)$$

Using AFM scans of graphene on ReS$_2$ we extract an average value for the bubble geometric parameters: $a = 74 \pm 2\ nm$, $h = 6 \pm 1\ nm$, and $d = 500 \pm 100\ nm$. For the water contact angles, we use $\theta_{graphene} = 64°$ [18] and $\theta_{ReS_2} = 60°$. We consider water contact angle for ReS$_2$ similar to that of other 2D materials [26,27]. Finally, we obtain the minimum migration temperature $T = 157°C$. This is consistent to measurements we performed by thermally annealing graphene/ReS$_2$ in a furnace.

To relate the laser power, the temperature in the bubble and the interface thermal conductance, we solved the steady state heat equation, following the procedures used in [28-32]. By equating this value with the minimum migration temperature obtained above, we extract the interface thermal conductance for graphene on ReS$_2$ to be g=0.725 MW/ (m$^2$K), a parameter that has not been reported thus far. We note that the value of interface thermal conductance varies depending on the 2D layers forming the interface. While the literature has still large variations for this parameter in 2D heterostructures, the value we found is close to reported experimental values for interfaces of MoS$_2$ with Au [33], and smaller than calculated for MoS$_2$ and SiO$_2$ [34,35] or graphene/BN [36]. This suggests that stronger adhesion leads to better heat conduction at interfaces.

The importance of reaching the minimum migration temperature is experimentally demonstrated in Figure 3, where it is apparent that a minimum laser power is required for flattening the graphene flake. A graphene/ReS$_2$ heterostructure (Figure 3(a)) is exposed to the laser at 4.35mW power. According to our model this corresponds to a bubble content temperature 120ºC, below the minimum migration temperature for this heterostructure. The result is presented in the AFM image of Figure 3(b), showing no change in the bubble distribution. We then increase the power to 8.7 mW (corresponding to 215ºC) and, as seen in Figure 3(c), we obtain larger regions of flat graphene. We note that we do not find the graphene to be damaged by this process, as verified by Raman spectroscopy (Figure 3(d)) which does not indicate the presence of a defect feature or changes in the typical graphene peaks.

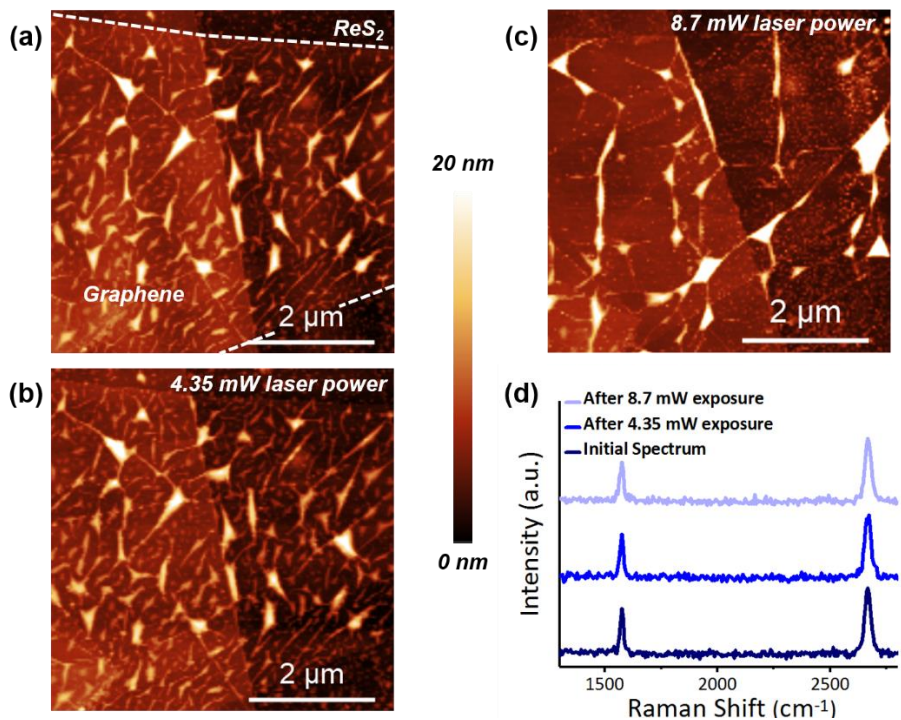

*Figure 3:* *Flattening requires sufficient laser power heating. **a**) AFM of a graphene flake supported by a ReS$_2$ substrate. **b**) AFM of the same graphene flake after laser exposure at 4.35 mW, indicating no bubble diffusion or morphological change. **c**) AFM of the same graphene flake after laser exposure at 8.7 mW, demonstrating that ~8mW is required to diffuse the transfer bubbles. **d**) Post exposure spectra showing no defect activated Raman peak.*

While for a given adhesion energy one needs an adequately high laser power to overcome it, the same laser power might not be sufficient for an interface with a higher adhesion energy. To demonstrate that, we consider a graphene/BN interface. As inferred from the aspect ratio of bubbles created at this interface ([18] and Figure 1(c)), the adhesion energy is larger and thus corresponds to a larger minimum migration temperature. In Figures 4(a) and 4(b) we show the results obtained when we applied our method to a heterostructure having graphene supported by thick (30-100nm) BN substrate. We expose these structures to identical laser rastering paths as the graphene on ReS$_2$ samples and observe that for the same laser power there is no bubble migration or coalescence. This finding is consistent with our model, as the adhesion energy at the graphene/hBN interface is larger than at the graphene/ReS$_2$ interface ($E_{adhesion\ G/hBN} = 126 mJ/m^2$ ) and at the same time, the losses to the substrate are larger, as the thermal conductance of graphene on hBN [36] ($g = 52.2 MW/m^2 K$) is greater than in the ReS$_2$ case. Using the model described above and the typical geometric parameters we obtained for bubbles at the graphene/hBN interface, we estimate that flattening graphene on hBN requires temperatures of T~324°C, consistent with values used for thermal annealing.

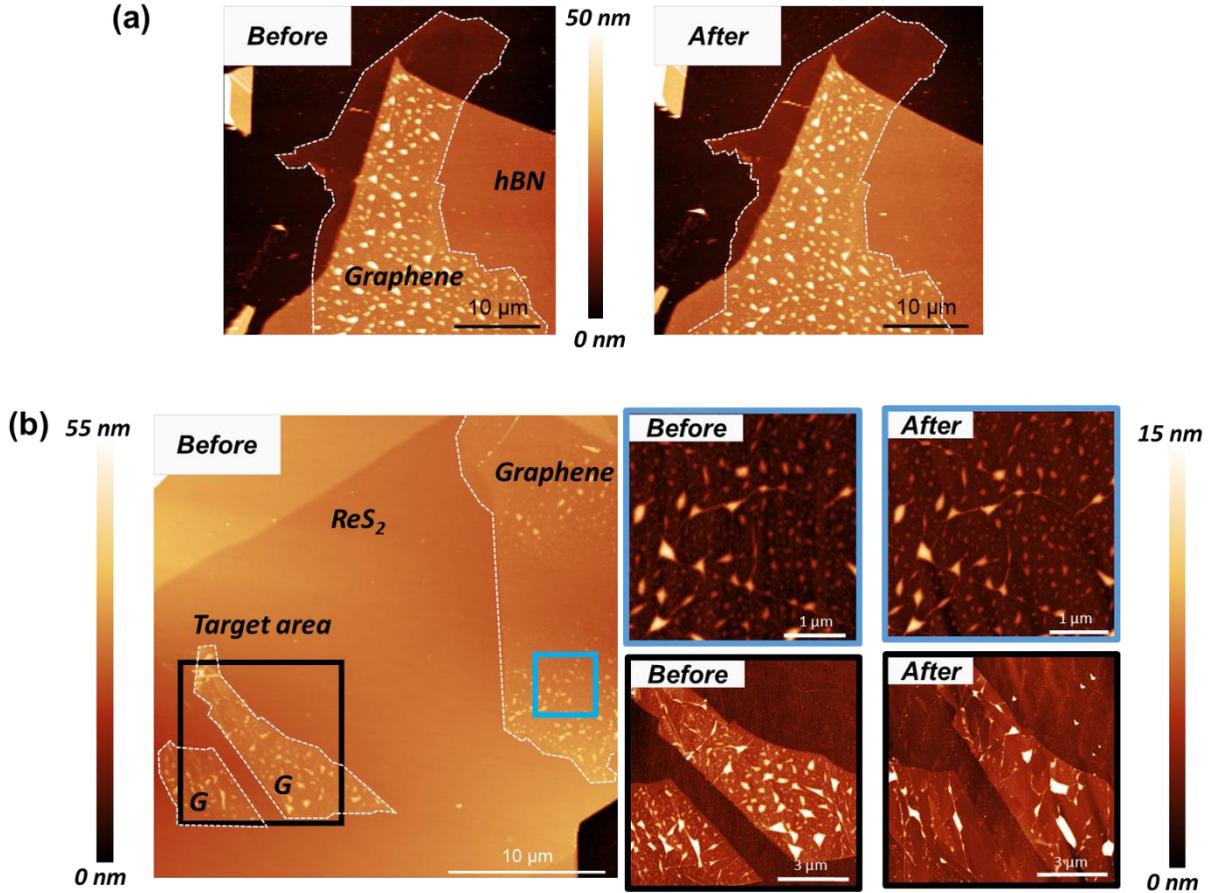

***Figure 4:** **a)** AFM of a graphene flake supported by hBN before and after an 8.7 mW laser exposure. **b)** AFM of separated graphene flakes supported by the same ReS$_2$ substrate before and after exposure to the target area at 6 mW. A high-resolution AFM topograph of the targeted flake (black box) shows substantial flattening whereas the unexposed flake (blue box) demonstrates no bubble changes.*

A great strength of the method we propose here is offering an alternative for a situation when the entire sample cannot be exposed to heat. Using the technique of localized thermal laser treatment, it is possible to flatten out specific and targeted sections of graphene supported by ReS$_2$ substrates by programming the laser path to only cover the desired spots. This feature is demonstrated in Figure 4(b) where two graphene flakes are placed on a ReS$_2$ substrate. We indicate the target area and as shown by the zoom-in topographs in the right side of Figure 4(b), we will only flatten the graphene flake in the desired region (black box) and not the on the different flake (blue box).

In conclusion, we present a method for eliminating bubbles at 2D materials interfaces based on localized thermal laser treatment. We discuss its range of applicability based on a model that considers the balance of heat loss at the interface and adhesion energy. By exemplifying its application to a graphene/ReS$_2$ interface we were also able to extract the interface thermal conductance for this system $g = 0.725 \frac{MW}{Km^2}$. The demonstrated control over the section to be flattened offers great advantage for optimizing custom device fabrication.

## ACKNOWLEDGMENTS

The authors acknowledge funding from the National Sciences and Engineering Research Council (NSERC) Discovery Grant RGPIN-2016-06717. We also acknowledge the support of the Natural Sciences and Engineering Research Council of Canada (NSERC) through Strategic Project STPGP 521420 (Quantum Circuits in 2D materials). Growth of hexagonal boron nitride crystals (K.W. and T.T.) was supported by the Elemental Strategy Initiative conducted by the MEXT, Japan and the CREST (JPMJCR15F3), JST. We thank Prof. Jean-Michel Menard and Prof. Rui Huang for useful discussions and Emmanuelle Launay for technical support.

## REFERENCES

[1]     P. Ajayan, P. Kim, and K. Banerjee, Physics Today **69**, 9 (2016).
[2]     A. K. Geim and I. V. Grigorieva, Nature **499**, 419 (2013).
[3]     K. S. Novoselov, A. Mishchenko, A. Carvalho, and A. H. Castro Neto, Science **353** (2016).
[4]     F. Pizzocchero, L. Gammelgaard, B. S. Jessen, J. M. Caridad, L. Wang, J. Hone, P. Bøggild, and T. J. Booth, Nature Communications **7**, 11894 (2016).
[5]     S. J. Haigh, A. Gholinia, R. Jalil, S. Romani, L. Britnell, D. C. Elias, K. S. Novoselov, L. A. Ponomarenko, A. K. Geim, and R. Gorbachev, Nature Materials **11**, 764 (2012).
[6]     A. V. Kretinin, Y. Cao, J. S. Tu, G. L. Yu, R. Jalil, K. S. Novoselov, S. J. Haigh, A. Gholinia, A. Mishchenko, M. Lozada, T. Georgiou, C. R. Woods, F. Withers, P. Blake, G. Eda, A. Wirsig, C. Hucho, K. Watanabe, T. Taniguchi, A. K. Geim, and R. V. Gorbachev, Nano Letters **14**, 3270 (2014).
[7]     A. S. Mayorov, R. V. Gorbachev, S. V. Morozov, L. Britnell, R. Jalil, L. A. Ponomarenko, P. Blake, K. S. Novoselov, K. Watanabe, T. Taniguchi, and A. K. Geim, Nano Letters **11**, 2396 (2011).
[8]     L. Wang, I. Meric, P. Y. Huang, Q. Gao, Y. Gao, H. Tran, T. Taniguchi, K. Watanabe, L. M. Campos, D. A. Muller, J. Guo, P. Kim, J. Hone, K. L. Shepard, and C. R. Dean, Science **342**, 614 (2013).
[9]     R. Frisenda, E. Navarro-Moratalla, P. Gant, D. Pérez De Lara, P. Jarillo-Herrero, R. V. Gorbachev, and A. Castellanos-Gomez, Chemical Society Reviews **47**, 53 (2018).
[10]    J. Boddison-Chouinard, R. Plumadore, and A. Luican-Mayer, JoVE, e59727 (2019).
[11]    E. Stolyarova, D. Stolyarov, K. Bolotin, S. Ryu, L. Liu, K. T. Rim, M. Klima, M. Hybertsen, I. Pogorelsky, I. Pavlishin, K. Kusche, J. Hone, P. Kim, H. L. Stormer, V. Yakimenko, and G. Flynn, Nano Letters **9**, 332 (2009).
[12]    T. Uwanno, Y. Hattori, T. Taniguchi, K. Watanabe, and K. Nagashio, 2D Materials **2**, 041002 (2015).
[13]    D. G. Purdie, N. M. Pugno, T. Taniguchi, K. Watanabe, A. C. Ferrari, and A. Lombardo, Nature Communications **9**, 5387 (2018).

[14] A. Jain, P. Bharadwaj, S. Heeg, M. Parzefall, T. Taniguchi, K. Watanabe, and L. Novotny, Nanotechnology **29**, 265203 (2018).
[15] M. R. Rosenberger, H.-J. Chuang, K. M. McCreary, A. T. Hanbicki, S. V. Sivaram, and B. T. Jonker, ACS Applied Materials & Interfaces **10**, 10379 (2018).
[16] H. Ghorbanfekr-Kalashami, K. S. Vasu, R. R. Nair, F. M. Peeters, and M. Neek-Amal, Nature Communications **8**, 15844 (2017).
[17] E. Khestanova, F. Guinea, L. Fumagalli, A. K. Geim, and I. V. Grigorieva, Nature Communications **7**, 12587 (2016).
[18] D. A. Sanchez, Z. Dai, P. Wang, A. Cantu-Chavez, C. J. Brennan, R. Huang, and N. Lu, Proceedings of the National Academy of Sciences **115**, 7884 (2018).
[19] D. Lloyd, X. Liu, N. Boddeti, L. Cantley, R. Long, M. L. Dunn, and J. S. Bunch, Nano Letters **17**, 5329 (2017).
[20] K. Yue, W. Gao, R. Huang, and K. M. Liechti, Journal of Applied Physics **112**, 083512 (2012).
[21] N. G. Boddeti, S. P. Koenig, R. Long, J. Xiao, J. S. Bunch, and M. L. Dunn, Journal of Applied Mechanics **80** (2013).
[22] T. Georgiou, L. Britnell, P. Blake, R. V. Gorbachev, A. Gholinia, A. K. Geim, C. Casiraghi, and K. S. Novoselov, Applied Physics Letters **99**, 093103 (2011).
[23] S. R. Na, J. W. Suk, R. S. Ruoff, R. Huang, and K. M. Liechti, ACS Nano **8**, 11234 (2014).
[24] J. S. Bunch and M. L. Dunn, Solid State Communications **152**, 1359 (2012).
[25] P. Bampoulis, V. J. Teernstra, D. Lohse, H. J. W. Zandvliet, and B. Poelsema, The Journal of Physical Chemistry C **120**, 27079 (2016).
[26] A. Kozbial, X. Gong, H. Liu, and L. Li, Langmuir **31**, 8429 (2015).
[27] Y. Wu, L. K. Wagner, and N. R. Aluru, The Journal of Chemical Physics **144**, 164118 (2016).
[28] W. Cai, A. L. Moore, Y. Zhu, X. Li, S. Chen, L. Shi, and R. S. Ruoff, Nano Letters **10**, 1645 (2010).
[29] C. Faugeras, B. Faugeras, M. Orlita, M. Potemski, R. R. Nair, and A. K. Geim, ACS Nano **4**, 1889 (2010).
[30] J.-U. Lee, D. Yoon, H. Kim, S. W. Lee, and H. Cheong, Physical Review B **83**, 081419 (2011).
[31] R. Yan, J. R. Simpson, S. Bertolazzi, J. Brivio, M. Watson, X. Wu, A. Kis, T. Luo, A. R. Hight Walker, and H. G. Xing, ACS Nano **8**, 986 (2014).
[32] A. A. Balandin, Nature Materials **10**, 569 (2011).
[33] X. Zhang, D. Sun, Y. Li, G.-H. Lee, X. Cui, D. Chenet, Y. You, T. F. Heinz, and J. C. Hone, ACS Applied Materials & Interfaces **7**, 25923 (2015).
[34] S. V. Suryavanshi, A. J. Gabourie, A. B. Farimani, and E. Pop, Journal of Applied Physics **126**, 055107 (2019).
[35] E. Yalon, Ö. B. Aslan, K. K. H. Smithe, C. J. McClellan, S. V. Suryavanshi, F. Xiong, A. Sood, C. M. Neumann, X. Xu, K. E. Goodson, T. F. Heinz, and E. Pop, ACS Applied Materials & Interfaces **9**, 43013 (2017).
[36] Y. Liu, Z.-Y. Ong, J. Wu, Y. Zhao, K. Watanabe, T. Taniguchi, D. Chi, G. Zhang, J. T. L. Thong, C.-W. Qiu, and K. Hippalgaonkar, Scientific Reports **7**, 43886 (2017).